\documentclass[11pt]{article}                     
\usepackage{graphicx}

\newcommand{\vs}{\vspace*{0.2 true cm}}

\newcommand{\Ma}{{\cal M}}

\newcommand{\0}{\phantom{0}}                      

\newcommand{\ADN} {{\it At. Nucl. Data Tables }}
\newcommand{\NIM} {{\it Nucl. Instrum. Meth. }}
\newcommand{\NPA} {{\it Nucl. Phys. A }}
\newcommand{\PR}  {{\it Phys. Rev. }}
\newcommand{\PRL} {{\it Phys. Rev. Lett. }}

\begin{document}
   \begin{center}
   ~ \vspace {3mm}
   {\Large \bf
       The History of Nuclidic Masses and of their Evaluation
                   } \\  \vspace {2mm}

   {\large Georges Audi}                   \vspace {2mm}

   {\small\it
       Centre de Spectrom\'etrie Nucl\'eaire et de
       Spectrom\'etrie de Masse, CSNSM, \\ IN2P3-CNRS, et UPS,
       B\^atiment 108, F-91405 Orsay Campus, France}  \vspace {5mm}

   {\footnotesize\it
   Contribution to the special issue of  \\
   the ``International Journal of Mass Spectrometry" (IJMS) \\
   in the honor of the 65th anniversary of J\"urgen Kluge's birthday \\
   (submitted December 24, 2005, resubmitted January 27, 2006) }       

   \end{center}

   \begin{abstract}
   \noindent
   This paper is centered on some historical aspects of nuclear masses,
and their relations to major discoveries.
   Besides nuclear reactions and decays, the heart of mass measurements
lies in mass spectrometry, the early history of which will be reviewed
first.
   I shall then give a short history of the mass unit which has not
always been defined as one twelfth of the carbon-12 mass.
   When combining inertial masses from mass spectrometry with energy
differences obtained in reactions and decays, the conversion factor
between the two is essential.
   The history of the evaluation of the nuclear masses (actually atomic
masses) is only slightly younger than that of the mass measurements
themselves.
   In their modern form, mass evaluations can be traced back to 1955.
   Prior to 1955, several tables were established, the oldest one in
1935.
   \vs

   \end{abstract}

   \section*{1 ~ The history of nuclear masses} \label{sect:hmas}

   The history of nuclear masses is almost as old as that of nuclear
physics itself.
   It started with the development of mass spectrography in the late
1910's\footnote{
      Writing about history is a particular exercise, not
      straightforward for scientists.
      Stating {\em ``The history of nuclear masses is as old as
      nuclear physics"} depends of course on definition.
      A.H.~Wapstra remarked that: {\em ``One could argue that it
      started in 1869 when Mendeleiev published the periodical
      system of elements, in which (average) atomic masses were
      the basis."}
      I hope that, in this historical sketch, I have not deviated
      too far from the truth.
   }.
   Mass spectrography itself was born in 1898 from the works of Wilhelm
Wien.
   He analyzed, with a magnet, the so-called `channel rays'\footnote{
      or `kanalstrahlen', the stream of positive ions formed from
      residual gases in cathode ray tubes.
   }
   discovered 12 years earlier by Eugen Goldstein.

   In the following, I shall give the important steps in the early
history of mass spectrometry with special focus on nuclear masses.
   The guideline will be given by the discoveries in physics that
thrived on them, rather than by the techniques or results for
themselves.

   \subsection*{First mass spectrographs}

\begin{figure}[!ht]   
   \begin{center}
   \includegraphics[width=9 cm]{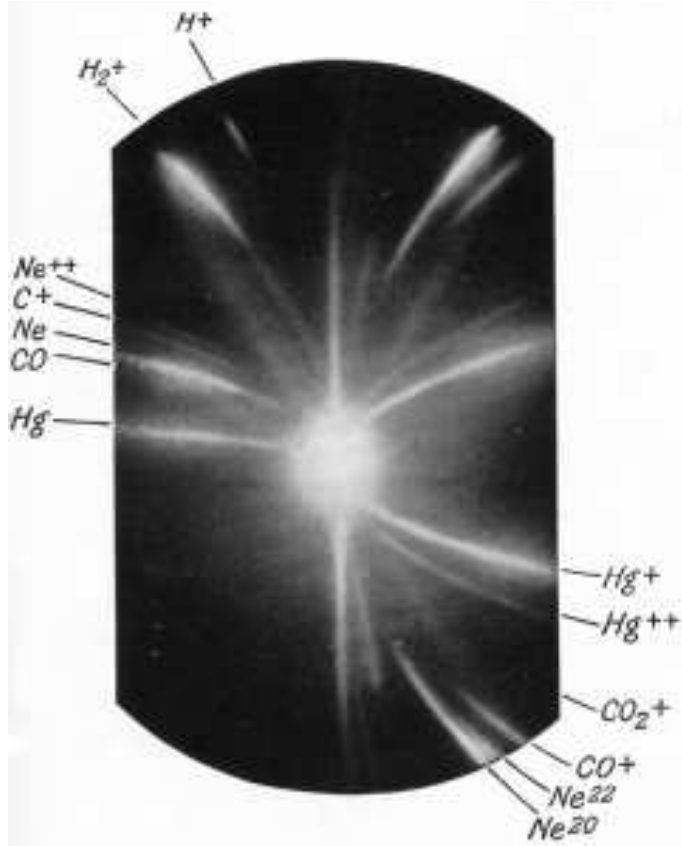}
   \caption[]{\footnotesize
      Photographic plate of the Thomson spectrograph.
      On this picture $\vec{B}$ and $\vec{E}$ are aligned along the
      horizontal axis $Ox$.
      Ions with the same mass will have positions $x=k_E\times q/{mv^2}$
      and $y=k_B\times q/{mv}$, then $x=\frac{k_E}{k_B^2}\times
      \frac{m}{q}\times y^2$ independently of their velocity $v$.
      They lie thus along a parabola.
      From Ref.\,\cite{42Aston}.
   }\label{fig:thomson}
   \end{center}
   \end{figure}

   In 1907, Joseph John Thomson\footnote{
      J.J.~Thomson was known already for his discovery of the
      electron in 1897, when he found that cathode rays were made
      of negative charged particles.
      He later built the plum-pudding model.
      An experiment of his former student, Ernest
      Rutherford, in 1911, showed that this model was not right.
   }
   built a spectrograph with aligned magnetic and electric fields, having
ions of the same species focused on the photographic plate along
parabolas, see Fig.\,\ref{fig:thomson}.
   The resolving power of this spectrograph was around $R=10-20$.
   In 1912, he obtained mass spectra of several gas compounds: N$_2$,
O$_2$, CO, CO$_2$,\ldots and was able to observe negatively-charged and
also multiply-charged ions.
   One year later (1913) he made one of the most important discoveries
in nuclear physics; he observed neon at two very different masses,
$A=20$ and $A=22$.
   This was the discovery of ``isotopism"\footnote{
      Frederick Soddy was the first to use the word ``isotope".
      He discovered, in 1910, that the average mass of natural lead
      (that we know today to be a mixture of four isotopes),
      and of lead obtained in the decay of uranium or of thorium
      differed beyond possible experimental uncertainties.
      He considered these ``isotopes" to be a peculiarity of
      radioactive materials:
      their nature was not understood at that time as due
      to different nuclidic species.
   }
   from direct observation of two different nuclidic species for the
same element.

   Then, a long series of improvements followed which increased the
resolving power and the sensitivity of the mass spectrographs and mass
spectrometers\footnote{
      The detector in a mass spectrograph is a photographic plate where
      ions of different masses strike at different locations.
      Whereas in mass spectrometers a spectrum is constructed by varying a
      parameter responsible for the acceleration or the deflection of the
      ions.
      The variations of ionic current at a fixed position is then
      recorded by means of an electrometer.
      The advantage of the photographic plate is its ability to
      simultaneously record lines corresponding to various ionic
      species.
      Also, in earlier times, the precision in the position of
      the lines was twice better.
      Only much later (see below) will the precision in the position of
      an electronically recorded peak gain several order of magnitudes.
      F.W.~Aston in Ref.\,\cite{42Aston}, p.\,38:
      {\em
      `` \ldots the word `mass spectrograph' has lost the original
      restrictions intended by the writer when he introduced it in 1920,
      and is now loosely applied to any method of positive ray analysis,
      even, by a quite unnecessary anachronism, to the parabola method.
      It is best restricted to those forms of apparatus capable of
      producing a focused mass spectrum of lines on a photographic plate.
      An apparatus in which the focused beam of rays is brought up to a
      fixed slit, and there detected and measured electrically is best
      termed a `mass spectrometer'.
      The first of these was devised by Dempster \ldots though the term
      was not introduced till much later."
      }
   }.

   The first of these improvements was introduced by Arthur Jeffrey
Dempster, at the University of Chicago, who, in 1918, built the first
mass spectrometer.
   Low-energy ions were accelerated to high energy (500 to 1750
volts) and deflected by a constant magnetic field.
   They were thus almost mono-energetic.
   A resolving power of $R=100$ was achieved by this
spectrometer\footnote{
      Dempster's second apparatus, in 1922, reached $R=160$ and
      precisions of $6\times10^{-4}$.
   }.

\begin{figure}[!ht]   
   \begin{center}
   \includegraphics[width=13 cm]{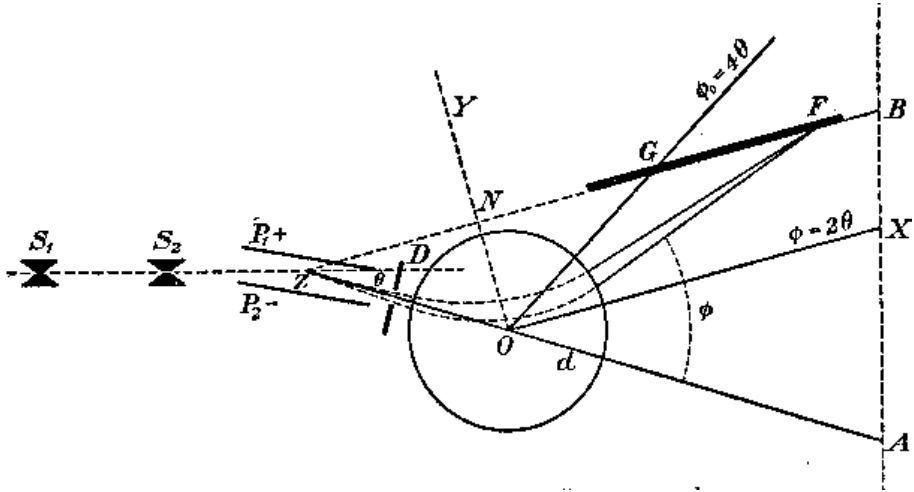}
   \caption[]{\footnotesize
   Diagram of Aston's first mass spectrograph (1919).
   The narrow slits $S_1$ and $S_2$ define a beam with very small
divergence.
   They are deflected and dispersed by the electric field between $P_1$
and $P_2$.
   The diaphragm $D$ selects ions in a small window in energy which
enter the magnetic field and are refocused at $F$ on the photographic
plate $GF$.
   From Ref.\,\cite{42Aston}.
   }\label{fig:aston}
   \end{center}
   \end{figure}

   In 1919 Francis William Aston, who was J.J.~Thomson's graduate student at Cambridge,
   built an instrument that was able to focus ions of the same species,
independently of their velocity spread (energy focusing).
   This increased the resolving power of his spectrograph up to $R=130$.
   He thus obtained relative precisions of $10^{-3}$ in mass
measurements\footnote{
      J.-L.~Costa, in 1925, built in Paris, an improved similar
      spectrograph and gained a factor two in resolving power
      and a precision of $3\times10^{-4}$.
   }
   with his first apparatus, see Fig.\,\ref{fig:aston}.

   With this limited precision he obtained two of the most remarkable
results.
   First, he was able to restore the ``whole number rule": all masses
(except hydrogen, see below) are whole numbers (which is true at this
level of $10^{-3}$), and a fractional `chemical' mass, like 35.5 for
chlorine, is in reality a mixture of the two isotopes at $A=35$ and 37
with ratios 3/4 and 1/4.
   The second remarkable result, and probably one of the most important
discovery for the story and the evolution of the Sun and the solar
system, is that hydrogen is an exception, with a mass of 1.008 (as
always, based on $^{16}$O$=16$).
   And this value {\em ``agrees with the value accepted by
chemists"} \cite{20Aston}.
   It was the theoretician and astronomer Arthur Stanley Eddington,
searching for a way out of the scientific crisis concerning the age of
the Sun, who immediately gave the answer.
   He calculated \cite{20Eddington} that the unaccounted energy
necessary for the extraordinary long lifespan of the Sun was due to
``sub-atomic energy":
   {\em
   ``There is sufficient {\rm (energy)} in the Sun to maintain its
output of heat for 15 billion years. \ldots
   Aston has further shown conclusively that the mass of the helium
atom is less than the sum of the masses of the 4 hydrogen atoms which
enter into it.\ldots
   There is a loss of mass in the synthesis amounting to about 1 part in
120, the atomic weight of hydrogen being 1.008 and that of helium just
4.\ldots
   We can therefore at once calculate the quantity of energy liberated
when helium is made out of hydrogen.
   If 5 per cent of a star's mass consists initially of hydrogen atoms,
which are gradually being combined to form more complex elements, the
total heat liberated will more than suffice for our demands, and we need
look no further for the source of a star's energy."
   }

\begin{figure}[!ht]   
   \begin{center}
   \includegraphics[width=14 cm]{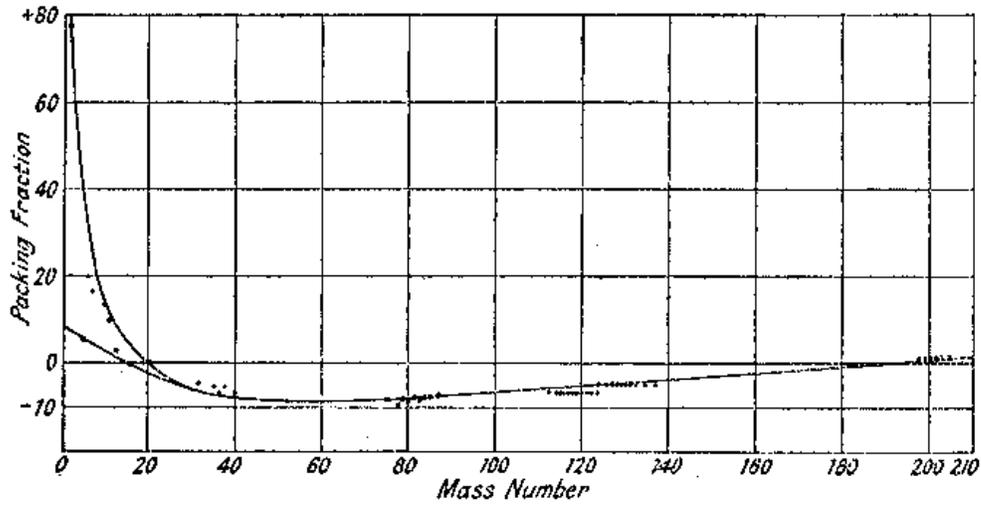}
   \caption[]{\footnotesize
   Aston's original packing fraction curve (1927).
   $P=10\,000\times\frac{\Ma-A}{A}$, where $\Ma$ is the atomic mass and
$A$ the mass number.
   From Ref.\,\cite{42Aston}.
   }\label{fig:aston-pack}
   \end{center}
   \end{figure}

   With his second spectrograph, built in 1925, Aston achieved a
resolving power of $R=600$ and could thus perform mass measurements with
a relative precision of $10^{-4}$.
   He could then observe that the actual masses were shifted by some
$8\times10^{-4}$ from the positions corresponding to full numbers,
discovering thus, in 1927, the ``mass defect"\footnote{
       Aston called the shift {\em ``packing fraction"} defined as the
       percentage of deviation of the masses from whole numbers
       (choosing $^{16}$O$=16$, a convention established by Aston and
       that lasted until 1960, see below) expressed in parts per 10\,000.
   },
   see Fig.\,\ref{fig:aston-pack}.
   (Aston built a third instrument in 1937 and reached $R=2000$).

\begin{figure}[!ht]   
   \begin{center}
   \includegraphics[width=10 cm]{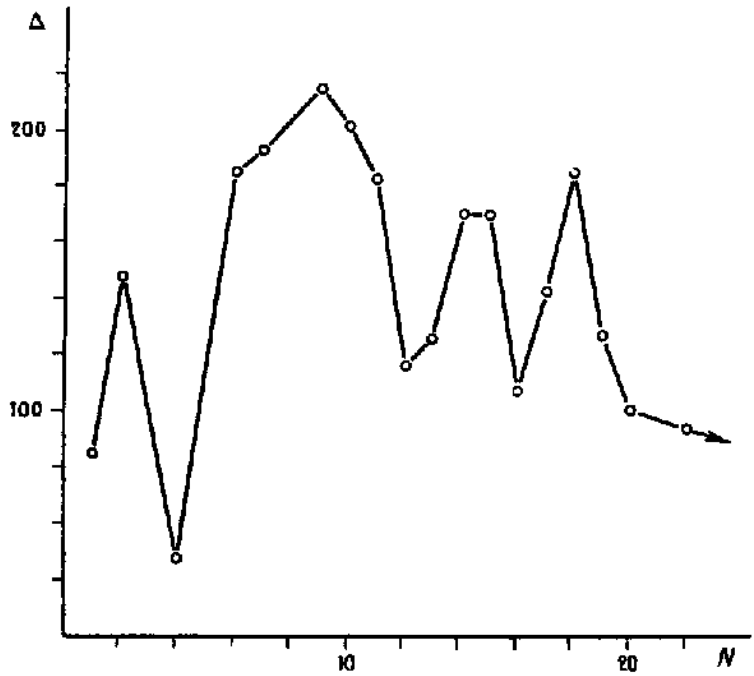}
   \caption[]{\footnotesize
   Plot of $\Delta=Mass-N\frac{35.9760}{36}$, where $N$ is here the mass
number.
   Clear minima of masses (maxima in binding energy) are evidenced at
mass numbers 4 and 16, i.e. for $^{4}$He and $^{16}$O.
   From Ref.\,\cite{33Elsasser}-a.
   }\label{fig:elsas}
   \end{center}
   \end{figure}

   The ensemble of masses obtained by Aston were determinant in the
discovery of closed shells by Walter M.~Elsasser in Paris in 1933:
   $Z=2$ and 8 (corresponding to mass numbers 4 and 16) in
Ref.\,\cite{33Elsasser}-a, see Fig.\,\ref{fig:elsas};
   $Z=20$, 28 and 50 in Ref.\,\cite{33Elsasser}-b.
   Elsasser also showed, in an independent study, that $Z=82$ and
$N=126$ correspond to shell closures based on the alpha decay energies,
see Ref.\,\cite{34Elsasser}.
   {\em ``One of the main nuclear features which led to the development
of the shell structure is the existence of what are usually called the
magic numbers.
   That such numbers exist was first remarked by Elsasser in 1933."}
\cite{63Mayer}.
   Strangely enough, the concept of shells was dropped by the
physicists of that time who were satisfied with the liquid drop model,
which seemed to efficiently describe fission.
   Shells were rediscovered 14 years later, in 1948, by Maria
Goeppert-Mayer \cite{48Mayer} when she examined a variety of
observables, including Elsasser's publications, looking closely at the
systematics of binding energies.
   The complete set of arguments, none of them individually conclusive
in her opinion, was convincing enough to claim the existence of ``magic
numbers"\footnote{
      It was actually Eugene Paul Wigner who coined the term
      {\em ``magic number"}.
      The physicists community at that time favored the liquid-drop
      model.
      {\em
      ``Wigner too believed in the liquid drop model, but he recognized,
      from the work of Maria Mayer, the very strong evidence for the
      closed shells.
      It seemed a little like magic to him, and that is how the
      words `Magic Numbers' were coined.",
      }
      said Steven A. Moszkowski, who was a student of Maria Goeppert-Mayer,
      in a talk presented at the APS meeting in Indianapolis, May 4, 1996.
      The rediscovery of ``magic numbers", lead M.~Goeppert-Mayer herself,
      and independently J.Hans~D.~Jensen in Europe, one year later
      in 1949, to the construction of the shell model with strong
      spin-orbit coupling, and to the Nobel prize they shared with
      Wigner in 1963.
   }
   \cite{48Mayer} at $N$ (or $Z$)$=20$, 50, 82 and 126.
   Magic numbers are much better demonstrated, nowadays, in the sudden
decrease in the separation energies\footnote{
      see, e.g. $S_{2n}$ and $S_{2p}$ graphs in \cite{Ame03a}, p.~542,
      and more graphs on the {\sc Amdc} web site \cite{amdcgr}.
   }
   after these numbers, similar to the ionization potential after
filling each atomic shell and responsible for the table of Mendeleiev.

   In 1932, Kenneth~T.~Bainbridge combined a Wien-type velocity filter
to a semi-circular magnet spectrograph and reached a resolving power of
$R=600$ and a $10^{-4}$ relative precision on masses.
   The measurements he performed allowed him to verify experimentally
the equivalence of mass and energy.
      F.W.~Aston in Ref.\,\cite{42Aston} p.\,85, says about K.T.~Bainbridge:
      {\em
      ``By establishing accurate comparisons of the masses of the light
      particles concerned in nuclear disintegrations, particularly
      that of $^7${\rm Li}, discovered by Cockcroft and Walton, he achieved a
      noteworthy triumph in the experimental proof of the fundamental
      theory of Einstein of the equivalence of mass and energy."
      }

   \subsection*{Double-focusing spectrographs}

\begin{figure}[!ht]   
   \begin{center}
   \includegraphics[width=14.5 cm]{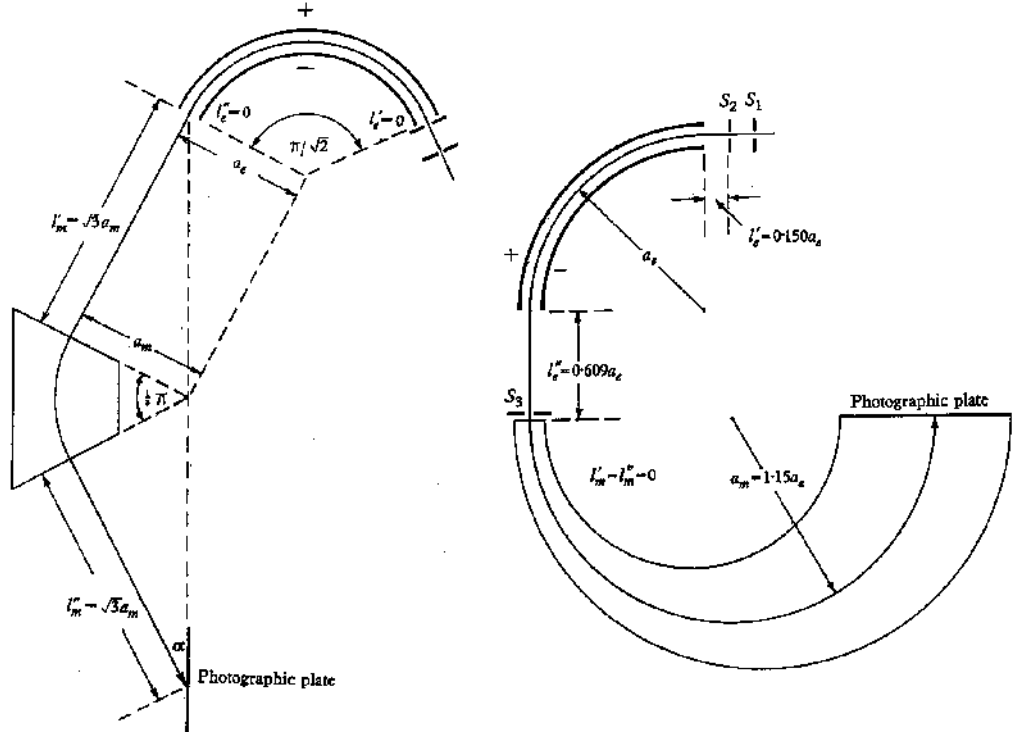}
   \caption[]{\footnotesize
       Two examples of double-focusing mass spectrographs:
       Bainbridge and Jordan's (left);
       and Dempster's (right).
       From Ref.\,\cite{58Duckworth}.
   }\label{fig:bjd}
   \end{center}
   \end{figure}

   The general idea then, in the early 30's, was to build a spectrograph
that would focus not only in velocity (energy) but also in direction:
the ``double-focusing" mass spectrograph.
   In a very short time interval, both concepts and designs flourished.
   Richard~F.K.~Herzog developed the theory of focusing in 1934.
   Then, simultaneously, and independently, three apparatuses were
built:
   one by Arthur J.~Dempster\footnote{
      R.C.~Barber comments \cite{05Barber}:
      {\em
      ``Following the first instruments, there were several that were
      developed with partial focusing based on simple geometric ideas.
      The development of the theory of focusing, by Herzog in 1934, gave
      rise to a generation of instruments in the late 30's that made use of
      the new insights."
      }
   }
   in Chicago, in 1935, yielding a resolving power of $R=3\,000$, see
Fig.\,\ref{fig:bjd}, right;
   another by Kenneth~T.~Bainbridge and Edward~B.~Jordan, at Harvard, in
1936, with a different geometry, achieving $R=10\,000$ and a mass
precision of $10^{-5}$, see Fig.\,\ref{fig:bjd}, left\footnote{
      four years later E.B.~Jordan reached $R=30\,000$.
   };
   the final by Josef~H.E.~Mattauch and Richard~F.K.~Herzog, in Vienna,
in 1936, with a resolving power of $R=6\,000$, for their first
spectrograph\footnote{
      several devices followed, built along the same line, yielding
      as much as $R=100\,000$.
   }.

\begin{figure}[!ht]   
   \begin{center}
   \includegraphics[width=12 cm]{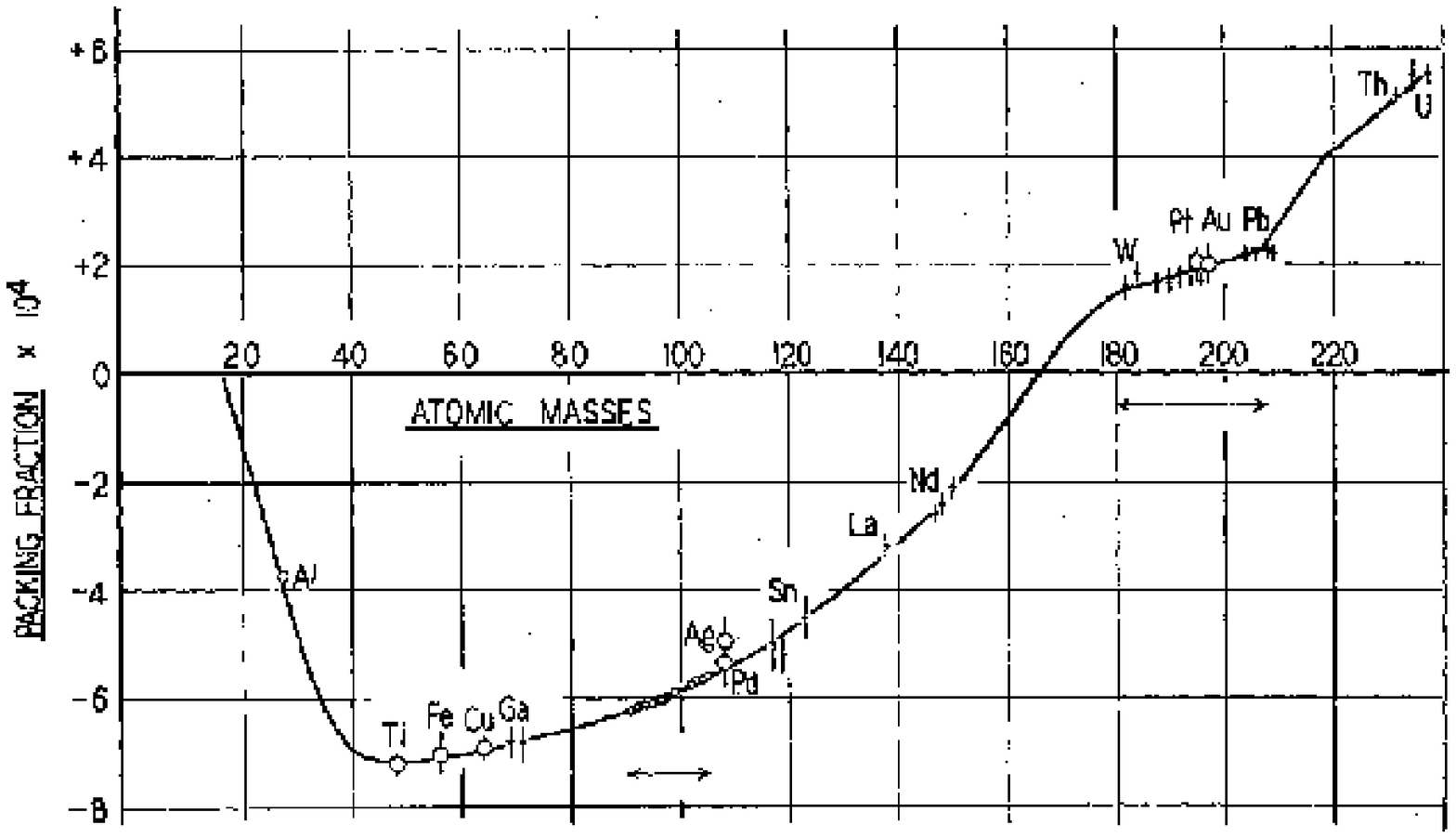}
   \caption[]{\footnotesize
      Dempster's packing fraction curve (1938).
      From Ref.\,\cite{38Dempster}.
   }\label{fig:demp-pack}
   \end{center}
   \end{figure}

\begin{figure}[!ht]   
   \begin{center}
   \includegraphics[width=13 cm]{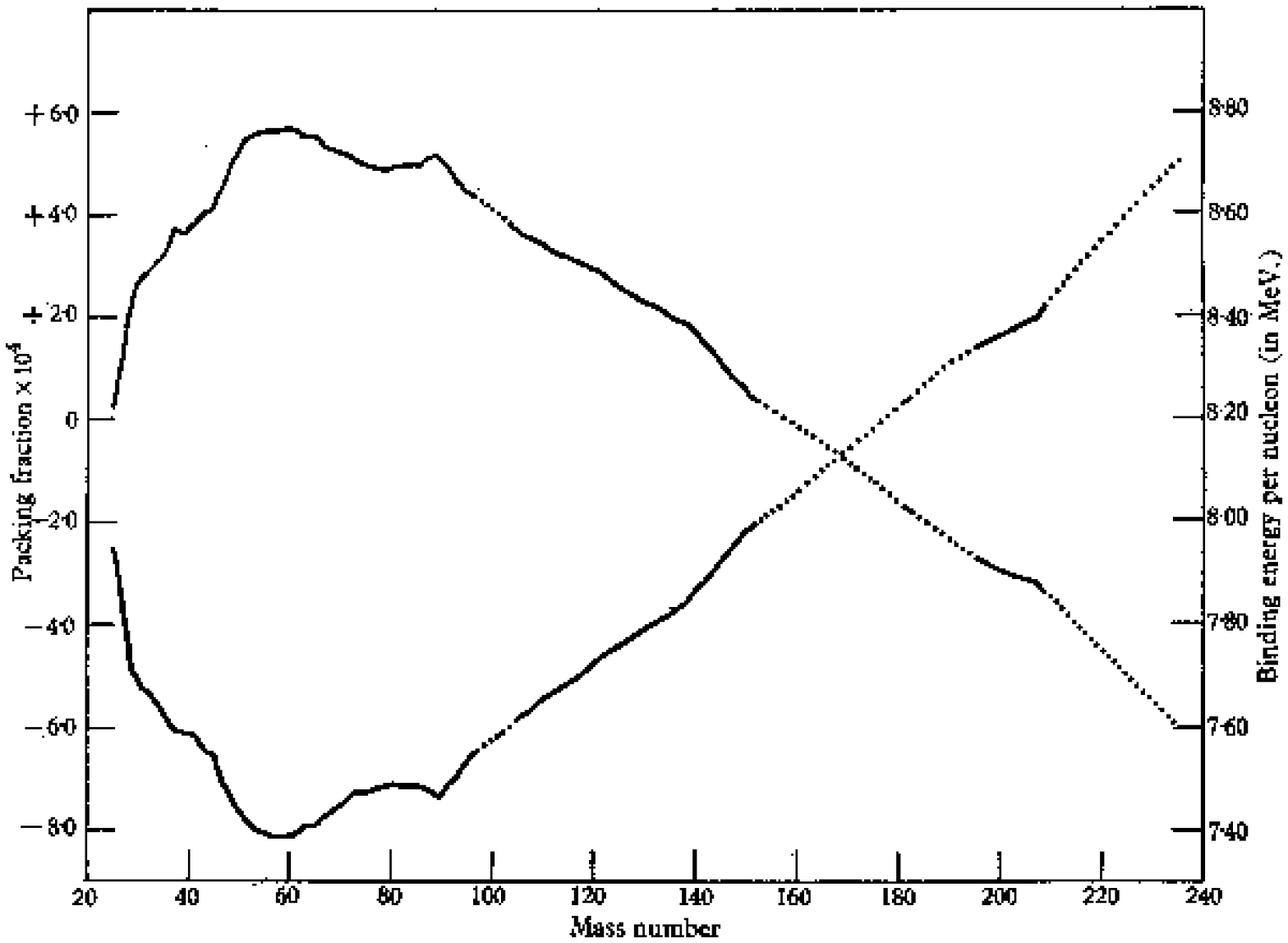}
   \caption[]{\footnotesize
      Duckworth's packing fraction curve (1958).
      The second curve with scale on the right is for the binding
      energy per nucleon.
      The now well known structures are clearly visible ($A=90$
      for magic $N=50$, $A=140$ for magic $N=82$, and $A=208$ for
      $^{208}_{82}$Pb$_{126}$).
      From Ref.\,\cite{58Duckworth}.
   }\label{fig:duck-pack}
   \end{center}
   \end{figure}

   Mass measurements with a precision of $10^{-5}$ were routinely
achieved then.
   The most remarkable result obtained by Dempster in 1938, only 3 years
after commissioning his spectrograph, was a greatly improved ``packing
fraction" curve, see Fig.\,\ref{fig:demp-pack}, that exhibited
structures not seen in Aston's (Fig.\,\ref{fig:aston-pack}).
   It is interesting to compare this 1938 curve to the ``modern" one of
Duckworth of 1958, Fig.\,\ref{fig:duck-pack}.

   Alfred~O.~Nier adopted, in the late 40's, the mass spectrometry's
detection technique
   (for the first time after the early Dempster's 1918 spectrometer, see
above).
   With his first double-focusing device, he obtained a resolving power
of $R=14\,000$; and, with his second enlarged version in 1956,
$R=75\,000$.

\begin{figure}[!ht]   
   \begin{center}
   \includegraphics[width=14.5 cm]{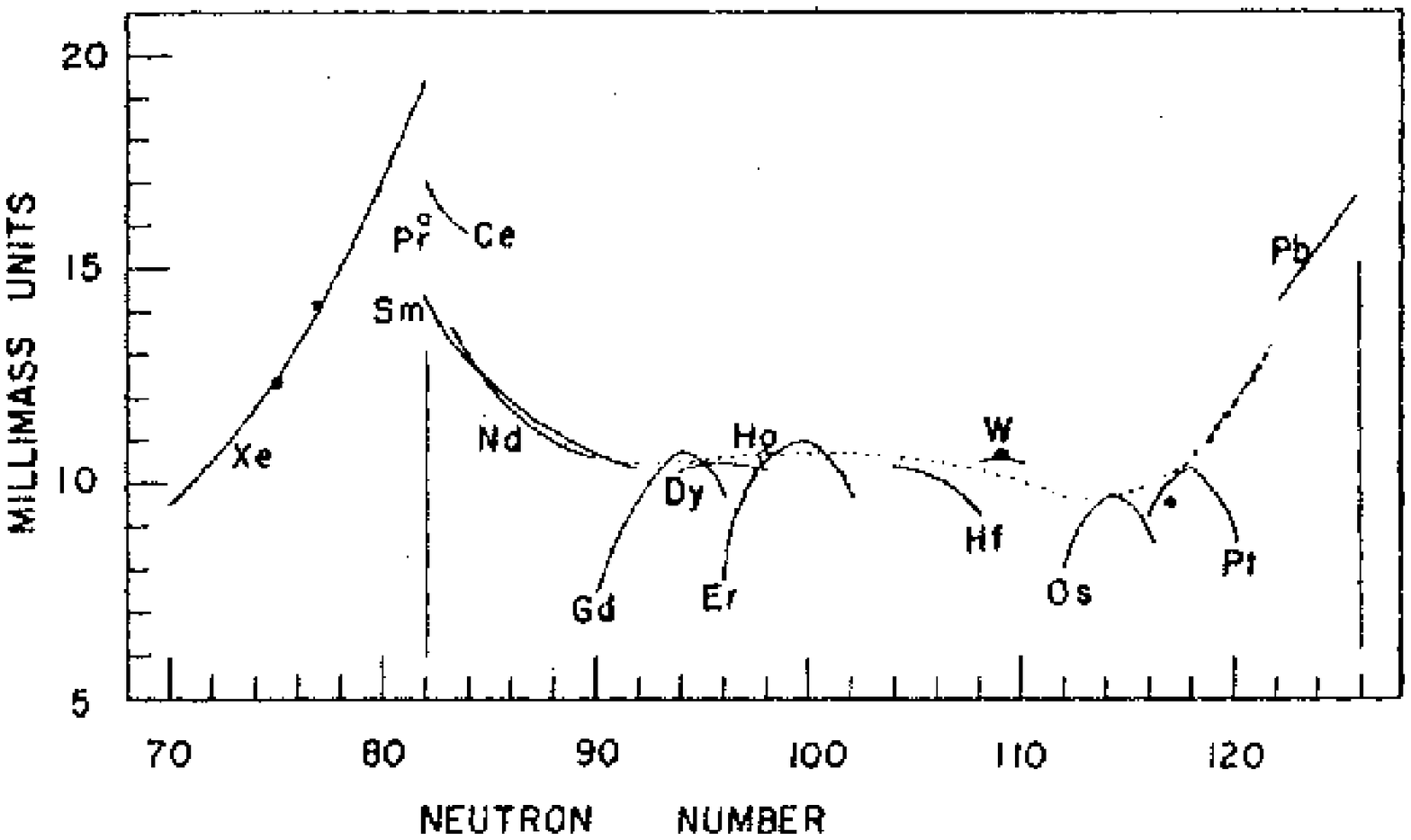}
   \caption[]{\footnotesize
   Plot of experimental deviation from a semi-empirical mass formula
   (expressed in milli-mass unit).
   The solid curves are the best fits for points belonging to even-$A$
even-$Z$ nuclei.
   From Ref.\,\cite{53Hogg}.
   }\label{fig:hogg}
   \end{center}
   \end{figure}

   Another important contribution to nuclear physics was brought by
Benjamin G.~Hogg and Henry~E.~Duckworth \cite{53Hogg}, in 1954, when
they discovered nuclear-shape deformation in the rare-earth region after
$N=90$, with a Dempster-type double-focusing mass spectrograph, see
Fig.\,\ref{fig:hogg}:
   {\em ``The extra stability in the heavier rare-earth region is not
adequately explained on a strict one-particle picture"}.
   They associated this extra stability with predicted ``mixing of
configurations".

   The interested reader will find more details, and also references to
the instruments mentioned above, and for which I gave no citation, in
three documents :
   \\ i) the book of Francis W.~Aston titled ``Mass Spectra and Isotopes"
\cite{42Aston};
   \\ ii) the book of Henry~E.~Duckworth titled ``Mass Spectroscopy"
\cite{58Duckworth}, and its update \cite{86Duckworth};
   \\ iii) the article ``Atomic Masses: Thomson to Ion Traps" by
Aaldert~H.~Wapstra \cite{95Wapstra}.

   \subsection*{Mass spectrometry of unstable nuclides - New spectrometers}

\begin{figure}[!ht]   
   \begin{center}
   \includegraphics[width=14 cm]{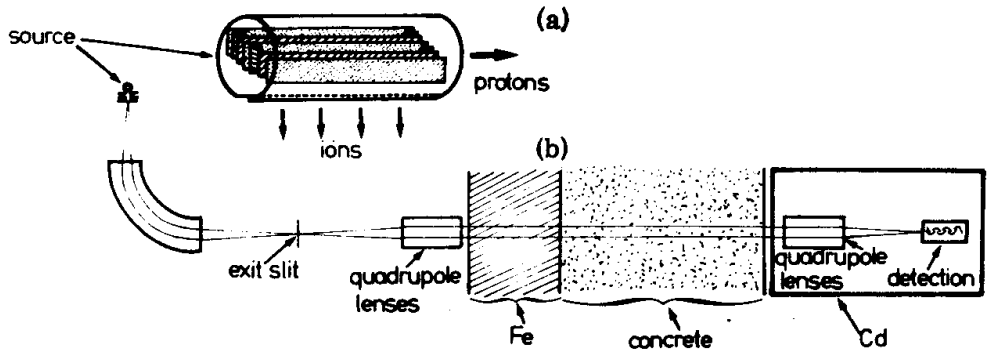}
   \caption[]{\footnotesize
   Klapisch and Thibault's mass spectrometer coupled to the PS
accelerator at {\sc Cern}.
   (a) enlargement of the target, made of graphite foils coated with
uranium, which is also the ion source of the spectrometer.
   (b) schematic lay-out of the spectrometer.
   Radioactive ions are produced by the proton beam impinging on uranium.
   They are focused in the ion source optics, enter the magnet, then
pass through the exit slit.
   They are then transported through iron, concrete and cadmium
shielding to a station where they are refocused and counted one by one
by a high gain ion multiplier.
   From Ref.\,\cite{73Klapisch}.
   }\label{fig:thib}
   \end{center}
   \end{figure}

\begin{figure}[!ht]   
   \begin{center}
   \includegraphics[width=14.5 cm]{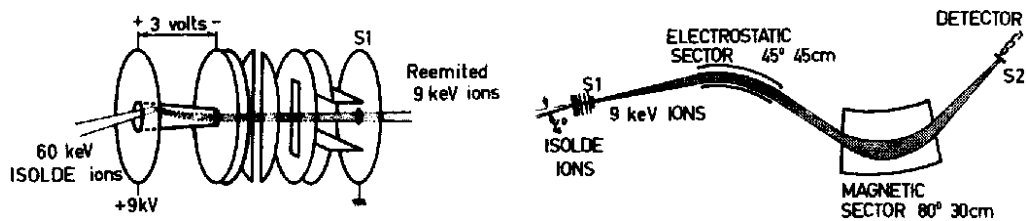}
   \caption[]{\footnotesize
   Diagram of Audi's double-focusing Mattauch-Herzog type
mass spectrometer coupled to the mass separator `on-line',
{\sc Isolde-II}.
   Left: the 60 keV {\sc Isolde} ions are stopped in the first atomic
layers inside a tantalum tube heated by a 3~volts DC current.
   The atoms diffuse out of the tantalum matrix, are reionized,
accelerated to 9 keV and focused on the entrance slit of the
spectrometer $S1$.
   Right: the radioactive ions travel through the spectrometer to the
exit slit $S2$ and are detected by an ion multiplier.
   From Ref.\,\cite{81Audi}.
   }\label{fig:audi}
   \end{center}
   \end{figure}

   In the early 1970's Robert Klapisch and Catherine Thibault
\cite{73Klapisch} coupled, for the first time, a classical
mass spectrometer to an accelerator (Fig.\,\ref{fig:thib}), the PS at
{\sc Cern}, to measure the masses of unstable species.
   They discovered that the magicity at $N=20$ disappeared for Na
isotopes.
   For my PhD thesis \cite{81Audi}, in 1981 in the same group, I coupled
a Mattauch-Herzog double-focusing spectrometer to {\sc Isolde-II}, also
at {\sc Cern} (Fig.\,\ref{fig:audi}), obtaining a resolving power of
$R=60\,000$.
   However, for the faint beams of radioactive species, I widened the
slits and achieved a typical value of $R=20\,000$ during operation.
   I found in the series of Rubidium isotopes a subshell closure at
$N=56$ and a deformation starting at $N=60$.

   In principle, history stops where the historian's own history starts.
   Even more so when the historian is an actor in the considered domain.
   But let me mention some of the most important steps that happened
since then.



   Around 1980, J\"urgen Kluge \cite{84Dabkiewicz} had the great idea to
exploit the fantastic resolving power of Penning traps\footnote{
         The ``Penning trap", first designed in 1949, is an instrument that
      combines an electric and a magnetic field in such a way that ions are
      trapped in a very small volume.
         Its development lead to one of the most drastic change in the
      landscape of mass spectrometry.
         The device received its name after Frans M.~Penning who, in 1936,
      `trapped' electrons in a magnetic field to increase their path in
      vacuum and thus increase the sensitivity of ionization vacuum gauges.
         (See the more detailed history of ion traps in
      Ref.\,\cite{95Holzscheiter}).
   }
   in order to perform nuclidic mass measurements.
   Before the end of that decade, he and his group effectively obtained
masses of unstable nuclear species with unprecedented precisions
\cite{90Stolzenberg}.

   In the late 1980's, Dave Pritchard \cite{89Cornell} at MIT built a
Penning trap for stable species, with which the incredible relative
precision of $10^{-10}$ for masses up to $A=40$ was obtained.
   Almost simultaneously Gerald Gabrielse \cite{86Gabrielse} built a
trap that he installed at {\sc Cern} close to an antiproton factory, to
compare the mass of the proton with that of the antiproton with a
precision of $9\times10^{-11}$, which is remarkable for a species so
difficult to isolate in our matter-dominated universe\footnote{
      The antiproton is expected to live as long as the proton.
      However, it annihilates in the presence of matter.
   }.
   He could thus prove the CPT conservation for the masses at this level
of precision.

   Today, Penning traps dominate the landscape of mass spectrometry,
not only for mass measurements, but in almost all fields using mass
spectrometers.
   Just to be complete, let us mention the other important developments
in mass spectrometry:
   the time-of-flight mass spectrometer of A.E.~Cameron and and D.F.~Eggers (1948),
   the radio-frequency mass spectrometer of Lincoln Smith (1960); and
   the so-called ``Schottky mass spectrometry" (1994), at {\sc Gsi}
using the {\sc Esr} storage ring as a spectrometer\footnote{
      named after Walter Schottky, 1886-1976, a German physicist who
      discovered the random noise due to the irregular arrival of electrons
      at the anode of thermionic tubes that is called ``shot noise".
        In Schottky mass spectrometry, ions circulate in a ring.
        At each turn, a detector consisting of pick-up electrodes
      (called ``Schottky" electrodes), records their time of passing.
        The rotation frequencies are thus determined.
        Therefore, strictly speaking, Schottky is not the name of a
      spectrograph, but that of a detection device, the spectrograph
      being the storage ring.
   }.


   In parallel to the development of mass spectrometry, important
development in ion sources extended the study to all types of elements.
   One can mention here the work of Alfred~O.~Nier, in 1940, on
electron-impact ion sources\footnote{
      {\em
      ``The construction of these instruments
      {\rm (spectrometers built in the mid 1950's)}
      was concurrent with new developments in ion optics, where the study
      of the effects of second order aberrations and of fringing fields
      were being pursued vigorously.
         ...
         The important development of vacuum and electronic technology during
      the WWII years led to enormous improvements in the post-war
      instruments."
      }
      \cite{05Barber}.
   }.

\begin{figure}[!ht]   
   \begin{center}
   \includegraphics[height=7.0 cm,bb= 80 34 574 560,clip,angle=270]{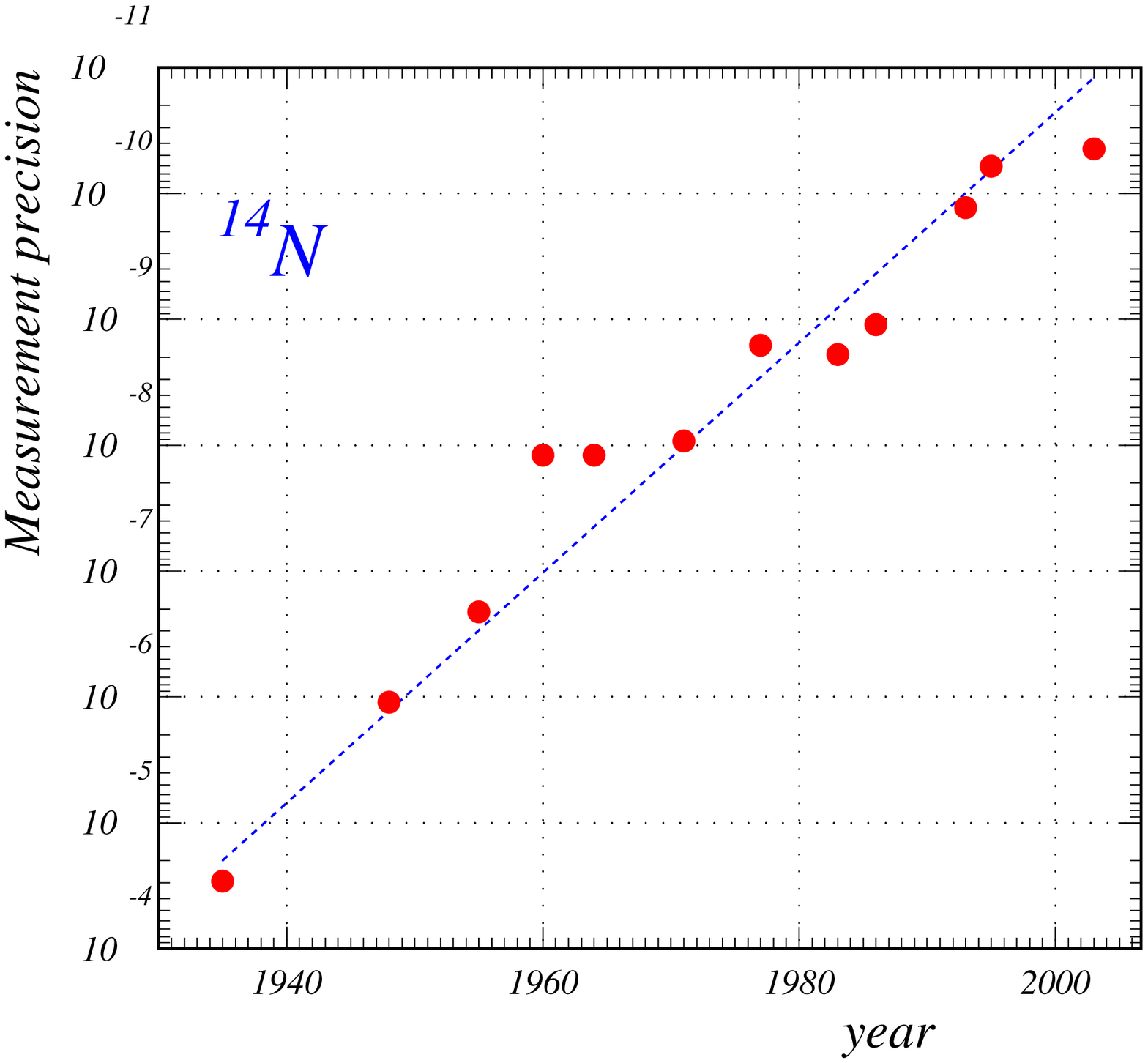}
   \includegraphics[height=7.0 cm,bb= 80 34 574 560,clip,angle=270]{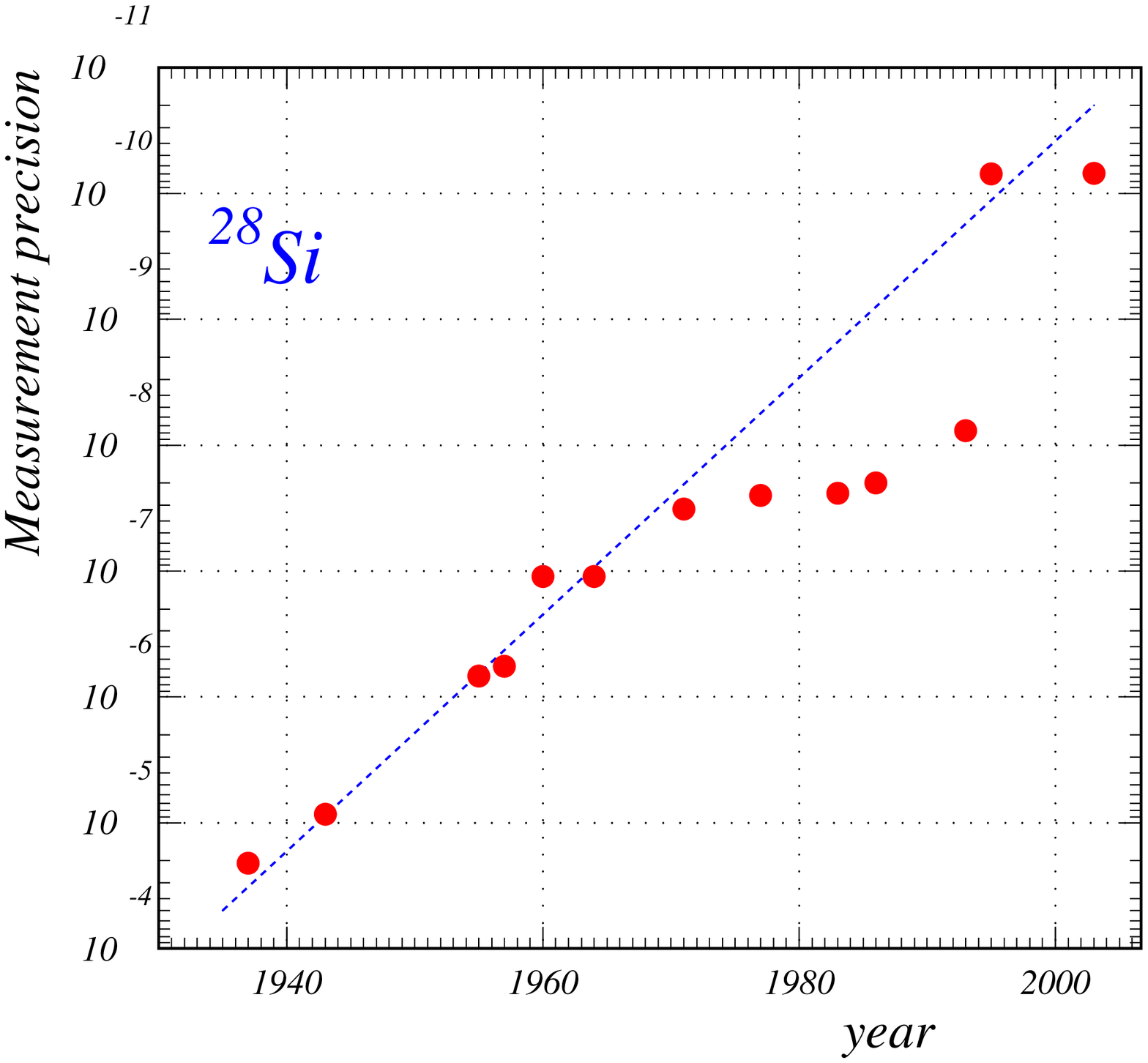}
   \caption[]{\footnotesize
   Evolution in the precision with which the masses of $^{14}$N and
$^{28}$Si have been known between 1937 and 2003.
   Our knowledge of these masses has increased by one order of magnitude
per decade.
   The values used to plot this figure are from mass tables,
respectively
   for year 1935 from Ref.\,\cite{35Bethe},
   1937 from Ref.\,\cite{37Livingston},
   1943 from \cite{43Flugge},
   1948 \cite{48Wapstra},
   1955 \cite{55Wapstra},
   1957 \cite{57Mattauch},
   1960 \cite{60Everling},
   1964 \cite{65Mattauch},
   1971 \cite{71Wapstra},
   1977 \cite{77Wapstra},
   1983 \cite{85Wapstra},
   1986 \cite{88Wapstra},
   1993 \cite{93Audi},
   1995 \cite{95Audi} and
   2003 \cite{Ame03a}.
   }\label{fig:28si}
   \end{center}
   \end{figure}

   Fig.\,\ref{fig:28si} illustrates the increase of precision obtained
in the last 70~years in the determination of the masses of $^{14}$N and
$^{28}$Si.
   Strikingly, on the average, one order of magnitude has been gained
every 10 years, from 400 and 600~$\mu$u respectively in 1937 to 0.6 and
2~nu in 2003.
   The precision in the mass of $^{28}$Si is important in view of the
redefinition of the last non-microscopic SI unit, the kilogram.
   One can remark a seemingly saturation of the precision in $^{28}$Si
after 1970 and for almost 20 years (the ``plateau" at 0.7~keV), and a
very rapid recovery in 1995 due to the MIT Penning trap work.
   Extrapolating the global tendency, one might expect a precision of
$10^{-11}$ or 0.2~nu in 2005, and that we will reach a precision of
$10^{-12}$ in 2015.
   As a matter of fact a paper just published \cite{2005Rainville} shows
that ratio of masses could be determined with a precision of
$7\times10^{-12}$, opening the possibilities to determine the mass of
$^{28}$Si with a precision close to this number.

   \section*{2 ~ A short history of the mass unit } \label{sect:hu}

   A mass measured by mass spectrometry is determined as an inertial
mass, from its movement characteristics in an electro-magnetic field.
   More exactly, the quantity measured ``directly" is the ratio of the
mass of the nuclide of interest to a well known mass.
   The result is then expressed in `unified atomic mass' (u)\footnote{
         Quite often people write erroneously `a.m.u' or `amu'
      instead of `u'.
   },
   or its sub-unit, $\mu$u.

   Mass measurements can also be obtained ``indirectly" as differences
in energy between neighboring nuclides, either by measuring a decay
energy or a reaction energy.
   An energy relation is thus established between the mass we want
to determine and a well known nuclidic mass.
   This energy relation is then expressed in electronvolts (eV).

   Two units are thus used in atomic mass measurements.
   We shall examine them separately and discuss how they are related.

   The mass unit is defined, since 1960, by 1\,u $= \Ma(^{12}$C$)/12$,
as one twelfth of the mass of one free atom of carbon-12 in its atomic
and nuclear ground-state.
   Before 1960, as Wapstra once told me, two mass units were defined:
   the physical one $\Ma(^{16}$O$)/16$,
   and the chemical one which represented one sixteenth of the average
mass of a standard mixture of the three stable isotopes of oxygen\footnote{
      R.C.~Barber's comment \cite{05Barber}:
      {\em
       ``The chemists used 1/16 of the mass of oxygen as the mass unit,
      beginning with very early work, back when they literally weighed the
      components in chemical reactions.
         As soon as Aston saw the evidence for isotopes of oxygen he realized
      that the definition, based on a `natural' abundance ratio for the
      three isotopes, was inherently imprecise.
         He defined all of his masses relative to $^{16}${\rm O}.
         This situation went on for decades.
         The chemists were untroubled by the slight difference, while the
      physicists were content with the Aston definition for a long time.
         However, with increasing precision, it was realized that mass
      spectroscopic comparisons were always referred to a `standard', known
      hydrocarbon molecule that always involved $^{12}${\rm C}.
         To convert from the defined standard $^{16}${\rm O} to $^{12}${\rm C},
      one had to study the $^{12}${\rm C}$^1${\rm H}$_4-^{16}${\rm O} doublet
      that was not particularly well known.
         If one changed to the $^{12}${\rm C} definition, there was an immediate
      `free' improvement in precision."
      }
   }.
   Physicists could not convince the chemists to drop their unit;
   {\em ``The change would mean millions of dollars in the sale of all
chemical substances"}, said the chemists, which is indeed true!
   Joseph~H.E.~Mattauch, the American chemist Truman~P.~Kohman and
Aaldert~H.~Wapstra \cite{58kohm} then calculated that, if
$\Ma(^{12}$C$)/12$ was chosen, the change would be ten times smaller for
chemists, and in the opposite direction\ldots
   That lead to unification;
   `u' stands therefore, officially, for `unified mass unit'!
   To be complete, let us mention that the chemical-mass spectrometry
community (e.g. bio-chemistry, polymer chemistry,\ldots) often use the
dalton\footnote{
      named after John Dalton, 1766-1844, a British scientist
      who first speculated that elements combine in proportions
      following simple laws, and was the first to create a table
      of (very approximate) atomic weights.
   }
   (symbol Da), which, whatever is claimed, serves actually to determine
the number of nucleons in a molecule, with not too much concern about
the exact value of the obtained mass compared to $^{12}$C.
   It is thus not strictly the same as `u'.
   As a matter of fact, some attempts were made to determine atomic
masses with equipment used in chemistry.
   However, the values obtained \cite{84Ni16,85Li02,90Me08} appeared
later to be at strong variance compared to modern results.

   The energy unit is the electronvolt.
   Until recently, the relative precision in $M-A$ expressed in keV was,
for several nuclides, less good than the same quantity expressed in mass
units.
   The choice of the volt for the energy unit (the electronvolt) is not
evident.
   One might expect the use of the {\em international} volt V, but one can
also choose the {\em standard} volt V$_{90}$ as maintained in national
laboratories for standards and defined by adopting an exact value for
the constant ($2e/h$) in the relation between frequency and voltage in
the Josephson effect.
   In the 1999 table of standards \cite{99mohr}:
$2e/h=483597.9$\,(exact)\,GHz/V$_{90}$.
   An analysis by E.Richard~Cohen and Aaldert H.~Wapstra \cite{83coh4}
showed that all precision measurements of reaction and decay energies
were calibrated in such a way that they can be more accurately expressed
in {\em standard} volt.
   Also, the precision in the conversion factor between mass unit and
{\em standard} volt, V$_{90}$, is more accurate than the conversion of
the mass unit to the {\em international} volt~V:
   \begin{eqnarray*}
   1 \, {\rm u} & = & 931\,494.009\,0   \pm  0.007\,1  \0 {\rm keV}_{90}  \\
   1 \, {\rm u} & = & 931\,494.013\,\0  \pm  0.037\,\0 \0 {\rm keV}
   \end{eqnarray*}

   This has not always been the case.
   In the early days of the evaluation of masses, two independent
evaluations and adjustments were often performed, separately for
reaction and decay data, and for mass spectrometric measurements.
   Comparing the two allowed one to derive a value for the conversion
factor, which could then be compared to the one derived by other, more
precise, methods \cite{99mohr}.
   The present computer program for the least-squares fit of the mass
adjustment still contains an option that allows this conversion factor
to be a free parameter.

   The reader will find more information on the energy unit, and also
some historical facts about the electronvolt, in the {\sc Ame2003}
\cite{Ame03}, page~134.

   \section*{3 ~ The history of the evaluation of atomic masses } \label{sect:hame}

   It was felt very early that establishing lists of properties for
nuclei was not only useful, but necessary.
   Several collections were thus published.
   Below is a list of the atomic mass compilations.
   The first one, to my knowledge, in which data from mass spectrometry
and nuclear reaction and decay data were combined, is the 1937 table of
Milton Stanley Livingston and Hans~Albrecht Bethe \cite{37Livingston}.

 {\footnotesize
 \begin{center}
 \begin{tabular}{llrl}
      1935 & H.~Bethe                        &  \cite{35Bethe}      & evaluation and table $^1n$-$^{17}$O
   \\ 1937 & M.S.~Livingston and H.A.~Bethe  &  \cite{37Livingston} & combined evaluation: energies + masses
   \\ 1943 & S.~Fl\"ugge and J.H.E.~Mattauch &  \cite{43Flugge}     &
   \\ 1944 & G.~Seaborg                      &  \cite{44Seaborg}    &
   \\ 1946 & G.~Seaborg                      &  \cite{46Seaborg}    & ``The Plutonium project table"
   \\ 1946 & J.~Suruque                      &  \cite{46Suruque}    &
   \\ 1948 & A.H.~Wapstra                    &  \cite{48Wapstra}    & ``Table of atomic nuclei"
   \\ 1953 & A.H.~Wapstra                    &  \cite{53Wapstra}    & $A>200$
 \end{tabular}
 \end{center}
 }

   In the early 1950's it was found that the many relations (direct and
indirect) overdetermined the mass value of many nuclides.
   Aaldert~H.~Wapstra
   established a procedure using a least-squares method to solve the
problem of overdetermination.
   One of the side-benefits of the overdetermination is to
get a check of the consistency among the various results.
   The first table of atomic masses using this method is
dated 1955 \cite{55Wapstra}.

   Since then, A.H.~Wapstra has carried on the evaluation of the
experimental masses of nuclei - to be more precise their atomic masses -
in what we call the Atomic Mass Evaluation ({\sc Ame}) with various
students or collaborators, until I joined him in 1981.
   We published together every 10 years since then (1983, 1993 and 2003)
a complete set of masses and of the data from which they are deduced.

   The list below gives the main ``modern" evaluations of atomic masses
following the general lines as first defined in Ref.\,\cite{55Wapstra},
well described in Ref.\,\cite{61Everling} and slightly refined since
then (see the most recent and most complete of those in the {\sc
Ame2003} \cite{Ame03}):

 {\footnotesize
 \begin{center}
 \begin{tabular}{llrl}
      1955 & A.H.~Wapstra and J.R.~Huizenga  &  \cite{55Wapstra}    & ``Isotopic masses"              
   \\ 1956 & J.~Mattauch {\it et al}         &  \cite{56Mattauch}   &  ``The masses of light nuclides 
   \\ 1957 & J.H.E.~Mattauch and F.~Everling &  \cite{57Mattauch}   &  ``Masses of atoms of $A<40$    
   \\ 1960 & F.~Everling {\it et al}         &  \cite{60Everling}   &  ``Relative nuclidic masses"    
   \\ 1962 & L.A.~K\"onig {\it et al}        &  \cite{62Konig}      &  ``1961 nuclidic mass table"
   \\ 1965 & J.H.E.~Mattauch {\it et al}     &  \cite{65Mattauch}   &  ``1964 atomic mass table"
   \\ 1971 & A.H.~Wapstra and M.B.~Gove      &  \cite{71Wapstra}    &  ``The 1971 atomic mass evaluation"
   \\ 1977 & A.H.~Wapstra and K.~Bos         &  \cite{77Wapstra}    &  ``The 1977 atomic mass evaluation"
   \\ 1985 & A.H.~Wapstra and G.~Audi        &  \cite{85Wapstra}    &  ``The 1983 atomic mass evaluation"
   \\ 1993 & G.~Audi and A.H.~Wapstra        &  \cite{93Audi}       &  ``The 1993 atomic mass evaluation"
   \\ 2003 & G.~Audi {\it et al}             &  \cite{Ame03a}       &  ``The {\sc Ame2003} atomic mass evaluation"
 \end{tabular}
 \end{center}
 }

   With the development of accelerators and the production of an
increasing number of unstable species, excited nuclear states were
increasingly populated.
   Those with half-lives long compared to typical electromagnetic
transitions (femtoseconds fs to picoseconds ps) started to play a {\it
r\^ole} in the measurement of decay energies.
   These long-lived excited states are called the excited
isomers\footnote{
      Here is another example of common misuse of terms.
      Strictly, we should distinguish ground-state isomer and excited
      isomers, as chemists distinguish left-handed and right-handed ones.
      Often people call `isomer' the excited one.
      (the Merriam-Webster says {\em ``isomerism: the relation of two
      or more nuclides with the same mass numbers and atomic numbers
      but different energy states and rates of radioactive decay"}).
   }.
   In many measurements of decay energies, it is not well established if
the emitting level from the mother nucleus is the ground-state or an
isomeric level.
   Often simultaneous measurement of the half-life or of the spin
(through the transition probabilities) removes the ambiguity.
   The problem, however, becomes worse in mass spectrometry where the
measurement may only yield one line for a mixture of closely lying
isomers.
   The interpretation is then difficult\footnote{
      Then, it is fully worthy to repeat the experiment with
      increased resolving power, even at the cost of lower
      counting rate and decreased precision.
      The final result will be more accurate.
      See the remarkable example of Hg isomers in Ref.\,\cite{01Sc41}.
   }.
   Only in favorable cases, corrections could be estimated.

   As a consequence, not only had the {\sc Ame} to handle several
isomers for each nuclidic species, but also one needed a unique and
consistent description of the various isomers that were involved in the
{\sc Ame}.
   Therefore the {\sc Nubase} database was created in 1993 and published
since then in 1997 and 2003 \cite{Nubase2003}, the latter in complete
synchronization with the {\sc Ame2003}.

   \section*{4 ~ Conclusion} \label{sect:conc}

   {\em ``The history of nuclear masses is almost as old as that of
nuclear physics itself."} was the first sentence of the introduction.
   The conclusion can complement this statement in that the history of
nuclear masses and the history of its most important contributor, namely
mass spectrometry, have continuously fed nuclear physics with major
discoveries.
   Each progress in building new spectrometers, increasing resolving
power or sensitivity or both, has led, as shown in this paper, to important
new physics.
   We have seen some of them:
   the nature of channel rays;
   the discovery of isotopism;
   the restoration of the whole number rule;
   the explanation of the age of the Sun;
   the discovery of mass defect and the experimental proof of the
equivalence of mass and energy;
   the discovery of the magic numbers;
   the discovery of deformations;
   the discovery of a subshell closure;
   the discovery that magic numbers might disappear;
   \ldots
   Much more could be said on recent breakthroughs in our understanding
of physics brought about by mass measurements, the history of which is
still to be written.

   In parallel to this history, and strongly related to it, is the
history of the atomic mass evaluation that establishes and ascertains
our confidence in the measured masses.

   \section*{Acknowledgements }

   I would like to thank Aaldert~H.~Wapstra who, since 1981, made me
discover the fascinating field of mass evaluation.
   Doing this job makes us see almost all aspects of the nuclear physics
problems and many also beyond.
   Often, we have the privilege to be the first ones to see new
phenomena.
   Therefore, thank you Aaldert for having let me enter this beautiful
domain.
   Now, we would both be so happy to have someone willing to share with
us the same passion.

   I want to thank J\"urgen Kluge, to whom this paper is dedicated in
honor of his 65th anniversary.
   J\"urgen invited me, in 1987, just after the first run with the
Penning trap at {\sc Isolde}, to join his team and enjoy the flourishing
program and the tremendous amount of data he and his group obtained with
this splendid spectrometer.

   I wish also to thank those who helped me in this work:
   first of all Lutz Schweikhard for asking me to write this piece of
history and for several very useful suggestions;
   and R.C.~Barber, F.~Everling, C.~Gaulard, D.~Lunney, C.~Thibault and
A.H.~Wapstra for reading, and making corrections and suggestions to
improve the manuscript.

\end{document}